\documentclass[12pt,preprint]{aastex}

\usepackage {graphicx}

\begin{document}

\title{Properties of Flares-Generated Seismic Waves on the Sun}
\author{A. G. Kosovichev}
\affil{W.W.Hansen Experimental Physics Laboratory, Stanford
University, Stanford, CA 94305, USA}

\begin{abstract}
The solar seismic waves excited by solar flares (``sunquakes'') are
observed as circular expanding waves on the Sun's surface. The first
sunquake was observed for a flare of July 9, 1996, from the Solar
and Heliospheric Observatory (SOHO) space mission. However, when the
new solar cycle started in 1997, the observations of solar flares
from SOHO did not show the seismic waves, similar to the 1996 event,
even for large X-class flares during the solar maximum in 2000-2002.
The first evidence of the seismic flare signal in this solar cycle
was obtained for the 2003 ``Halloween'' events, through acoustic
``egression power'' by Donea and Lindsey. After these several other
strong sunquakes have been observed. Here, I present a detailed
analysis of the basic properties of the helioseismic waves generated
by three solar flares in 2003-2005. For two of these flares, X17
flare of October 28, 2003, and X1.2 flare of January 15, 2005, the
helioseismology observations are compared with simultaneous
observations of flare X-ray fluxes measured from the RHESSI
satellite. These observations show a close association between the
flare seismic waves and the hard X-ray source, indicating that
high-energy electrons accelerated during the flare impulsive phase
produced strong compression waves in the photosphere, causing the
sunquake. The results also reveal new physical properties such as
strong anisotropy of the seismic waves, the amplitude of which
varies significantly with the direction of propagation. The waves
travel through surrounding sunspot regions to large distances, up to
120 Mm, without significant decay. These observations open new
perspectives for helioseismic diagnostics of flaring active regions
on the Sun and for understanding the mechanisms of the energy
release and transport in solar flares.
\end{abstract}

\section{Introduction}

 It was suggested long ago \citep{Wolff1972} that
solar flares, giant explosions on the Sun, may cause acoustic waves
traveling through the Sun's interior, similar to the seismic waves
on the Earth. Because the sound speed increases with depth the waves
are reflected in the deep layers of the Sun and appear back on the
surface, forming expanding rings of the surface displacement.
Theoretical modeling \citep{Kosovichev1995} predicted that the speed
of the expanding seismic waves increases with distance because the
distant waves propagate into the deeper interior where the sound
speed is higher. First observations of the seismic waves caused by
the X2.6 flare of July 9, 1996 \citep{Kosovichev1998}, proved these
predictions. These observations also showed that the source of the
seismic response was a strong shock-like compression wave
propagating downwards in the photosphere. This wave was observed
immediately after the hard X-ray impulse which produced by
high-energy electrons hitting the low atmosphere. This led to a
suggestion that the seismic response can be explained in terms of
so-called ``thick-target'' models. In these models, a beam of
high-energy is related to heating of the solar chromosphere,
resulting in evaporation of the upper chromosphere and a strong
compression of the lower chromosphere \citep[e.g.][]{Kostiuk1975,
Livshits1981, Fisher1985, Kosovichev1986}. This high-pressure
compression produces a downward propagating shock wave
\citep{Kosovichev1986} that hits the solar surface and causes
sunquakes. This shock observed in SOHO/MDI Dopplergrams as a
localized large-amplitude velocity impulse of about 1 km/s or
stronger represents the initial hydrodynamic impact resulting in the
seismic response. In addition, \citet{Kosovichev1998a} found that
the seismic wave was anisotropic, with a significant quadrupole
component.

 The following observations of solar flares made
by the Michelson Doppler Imager (MDI) instrument on the NASA-ESA
mission SOHO did not show noticeable sunquake signals even for
strong X-class flares. This search was carried out by calculating an
``egression'' power for high-frequency acoustic waves during the
flares \citep{Donea1999}. It became clear that sunquakes are a
rather rare phenomenon on the Sun, which occurs only under some
special conditions. Surprisingly, seven years later several flares
did show strong ``egression'' signals indicating new potential
sunquakes \citep{Donea2005} (for a list see
http://www.maths.monash.edu.au/\~\,adonea). It is interesting to
note that the flare of July 9, 1996, was the last strong of the
previous solar activity cycle, and the new strong sunquake events
are observed in the declining phase of the current activity cycle
after the maximum of 2000-2001. It appears that during the rising
phase of the solar cycle and during its peak the solar flares are
rather a ``superficial'' coronal phenomenon not affecting much the
solar surface and interior. This could happen if the topology of
magnetic field of solar active regions which produce flares changes
in such a way that the magnetic energy is released at lower
altitudes in the declining phase of the solar cycle than in the
rising and maximum phase. Here, I present analysis of new
observations of the seismic response to solar flares from the SOHO
and RHESSI space observatories, which show that the sunquakes were
indeed caused by the hydrodynamic impacts of high-energy electrons
accelerated in solar flares confirming the initial result of
\citep{Kosovichev1998}, and determine basic properties of the
flare-generated seismic waves by investigating their time-distance
characteristics.

\section{Results of analysis of SOHO/MDI and RHESSI data}

The MDI instrument on SOHO measures motions of the solar surface
through the Doppler shift of a photospheric absorption line Ni I
6768 A. The measurements provide images of the line-of-sight
velocity of the Sun's surface every minute with the spatial
resolution 2 arcsec per pixel. Examples of the MDI Dopplergrams
obtained during the sunquake events are shown in the two right
columns in Figure 1 (grey semitransparent images overlaying color
images of sunspots). There are several types of motions on the solar
surface, which contribute to the MDI signal. The largest
contributions of about 500 m/s come from the solar convection and
stochastic 5-min oscillations excited by convection (they form the
noisy granular-like pattern in Fig.1). The amplitude of the
flare-generated seismic waves (ring-like features identified in the
middle column of Fig.1) rarely exceeds 100 m/s. Thus, because of the
strong stochastic motions in the background, these waves are
difficult to detect. However, these waves form an almost
circular-shape expanding ring, velocity of which is determined by
the sound speed inside the Sun and can be calculated from solar
models. This property is used to extract the seismic response signal
from the noisy data. Because the waves are close to circular the
Dopplergrams can be averaged over a range of the azimuthal angle
around central points of the initial flare impact. These centers are
identified during the flare impulsive phase as strong localized
rapidly varying velocity perturbations of about 1 km/s (light and
dark features in left column of Fig.1). The azimuthally averaged
Dopplergrams are plotted as time-distance diagrams (right columns of
Fig.1; the averaging angular range in the polar coordinates in
indicated at the top), in which the seismic wave forms a continuous
ridge corresponding the time-distance relation for acoustic
propagating through the solar interior. The slope of this ridge is
decreasing with distance, meaning that the waves accelerate. This
happens because the seismic waves observed at longer distances
travel through the deeper interior of the Sun where the sound speed
is higher because of higher plasma temperature. Typically, the ring
speed changes from 10 km/s to 100 km/s. In the ``egression power''
method \citep{Donea1999} the wave signal is integrated along the
time-distance ridge, thus giving the total average of the seismic
signal power for specific central points. The egression power can be
calculated for the whole Dopplergram revealing places of potential
sunquakes. This method is useful as a search tool, but it does not
provide characteristics of the seismic waves. Because of the high
solar noise, the seismic waves are not easily seen on individual
Dopplergrams. They are much easier recognized in Dopplergram movies
as expanding circular wave fronts. The typical oscillation frequency
of the flare waves is higher than the mean frequency of the
background fluctuations (4-5 mHz vs. 3 mHz). Therefore, frequency
filtering centered at 5 or 6 mHz helps to increase the
signal-to-noise ratio. In most cases, a frequency filter centered at
6 mHz with the width of 2 mHz is used, and, in addition, the
difference filter for consecutive images is applied.

Localized Doppler perturbations during the flare impulsive phase,
similar to shown in the Fig.1 (left panels) and presumably
associated with precipitation of high-energy particles are commonly
observed. Therefore, that one might expect that seismic waves are
excited in most flares that affect the photosphere. However, in most
cases the flare hydrodynamic impact in the photosphere and, thus,
the seismic response appear to be weak. The main purpose of this
paper is to investigate properties of strong seismic waves when
their wave fronts can be observed explicitly in Dopplergrams and
time-distance diagrams. A list of 6 flares with such strong seismic
waves, observed from SOHO/MDI between 1996 and 2005 in given in
Table 1.

Figure 1 presents results for three strongest events so far,
observed on 10/28/2003, 07/16/2004 and 01/15/2005. The first flare
of October 28, 2003, was one of the strongest ever observed, having
the soft X-ray class X17. It is interesting that the two other
flares had much weaker soft X-ray class, but produced higher
amplitude seismic waves than this one. The analysis of these
observations reveals new interesting features of the seismic
response: 1) flares can produce multiple sunquakes almost
simultaneously originating from separate positions (as also found by
\citet{Donea2005} for the 10/28/2003 flare); 2) the seismic waves
are highly anisotropic, their amplitude can vary significantly with
angle; 3) the strongest amplitude is commonly observed in the same
direction as the direction of motion of flare ribbons; 4) the wave
fronts in most cases have elliptical shape, originating from
elongated in one direction initial impulse; 5) the centers of the
expanding waves coincide very well with the places of hydrodynamic
impacts in MDI Dopplergrams \citep[confirming the initial
observation of][]{Kosovichev1998}, however, not all impact sources
produce strong seismic waves; 6) the seismic waves are usually first
observed 15-20 min after the initial impact, and reach the highest
amplitude 20-30 min after the flare; 7) the seismic waves can travel
to large distances exceeding 120 Mm, but, in some cases, decay more
rapidly; 8) the fronts of acoustic seismic waves propagate through
sunspots without much distortion and significant decay, thus showing
no evidence for conversion into other types of MHD waves; 9) the
time-distance diagrams for the waves propagating in sunspot regions
show only small deviations of the order of 2-3 min from the wave
travel times of the quiet Sun; these variations are consistent with
the travel time measurements obtained by time-distance
helioseismology using the cross-covariance function for random waves
\citep{Duvall1997, Kosovichev2000}.

For two of these flares, X17 of October 28, 2003, and January 15,
2005, X-ray data are available for analysis. The RHESSI image
reconstruction software was used to obtain locations of the X-ray
sources in these flares and compare with the MDI Doppler
measurements of the hydrodynamic impulses and seismic responses.
Figure 2 shows a white-light image of the flaring active region
(NOAA 10696) and the superimposed images of the Doppler signal at
the impulsive phase, 11:06 UT, (blue and yellow spots show up and
down photospheric motions with variations in the MDI signal stronger
than 1 km/s), positions of three wave fronts at 11:37 UT, and also
locations of the hard X-ray (50-100 keV) sources (yellow circles) at
11:06 UT, and 2.2 MeV gamma-ray sources (green circles) found by
\citet{Hurford2004} (averaged for the whole flare duration).

Evidently, the X-ray and gamma-ray source are very close to the
positions of the seismic sources, but there was no gamma-ray
emission near source 3. Also, the gamma emission was not detected
for other seismic events. This leads to the conclusion that the
origin of the seismic response is the hydrodynamic impact (shock),
which is observed in the Doppler signals at 11:06 UT and shows the
best correspondence to the central positions of the wave fronts,
contrary to the suggestion of \citet{Donea2005} that photospheric
heating by high-energy protons is likely to be a major factor. This
was verified by calculating the time-distance diagrams for various
central positions and various angular sectors. When the central
position of a time-distance diagram deviates from the seismic source
position this deviation is immediately seen in the diagram as an
off-set of the time-distance ridge. This approach provides effective
source positions for complicated and distributed Doppler signals.

The flare of January 15, 2005, of moderate X-ray class, X1.2, but it
produced the strongest seismic wave observed so far by SOHO (Fig.1,
bottom row). Its amplitude exceeded 100 m/s. This wave had an
elliptical shape with the major axis in the SE-NW direction. The
elliptical shape corresponds very well to the linear shape of the
seismic source extended in this case along the magnetic neutral
line. This is illustrated in Figure 3. The left panel shows the
grey-scale map the Dopplergram difference at 0:40 UT, in which the
long white feature near the center corresponds to strong downflows
at the seismic source, and an image of the hard X-ray source (color
spot). The right panel shows the corresponding MDI magnetogram and
an image of the soft X-ray emission (in gray) and contour line of
the hard X-ray source. Evidently, the region of the hydrodynamic
impact was located just below the hard X-ray source, which was at a
footpoint of the soft X-ray loop. Figure 4 illustrates this sequence
of events for the January 15, 2005, flare, from top to bottom.

The high-energy electrons accelerated in the flare (presumably, high
in the corona) produced hard X-ray impulse in the lower atmosphere
and generated downward propagating shocks which hit the photosphere
and generated the seismic waves. This picture corresponds very well
to the standard thick target model of solar flares
\citep{Svestka1970} and the models of the hydrodynamic response
\citep[e.g.][]{Kostiuk1975, Livshits1981, Fisher1985, Kosovichev1986}.
The soft X-ray image indicates this flare was rather compact. One
may suggest that the seismic response can be particularly strong in
the case of a compact solar flare, but this needs to be confirmed by
further observations.

\section{Discussion}

The new observations from SOHO and RHESSI provide unique information
about the interaction of the high-energy particles accelerated in
solar flares with solar plasma and the dynamics of the solar
atmosphere during solar flares. These data also provide unique
information about the interaction of acoustic MHD waves with
sunspots, showing explicitly propagation of wave fronts through
sunspot regions. This opens opportunity for developing new methods
of helioseismology analysis of flaring active regions, similar to
the methods of Earth-quake seismology.

\clearpage

\begin{figure}
\centerline{\includegraphics[scale=0.8]{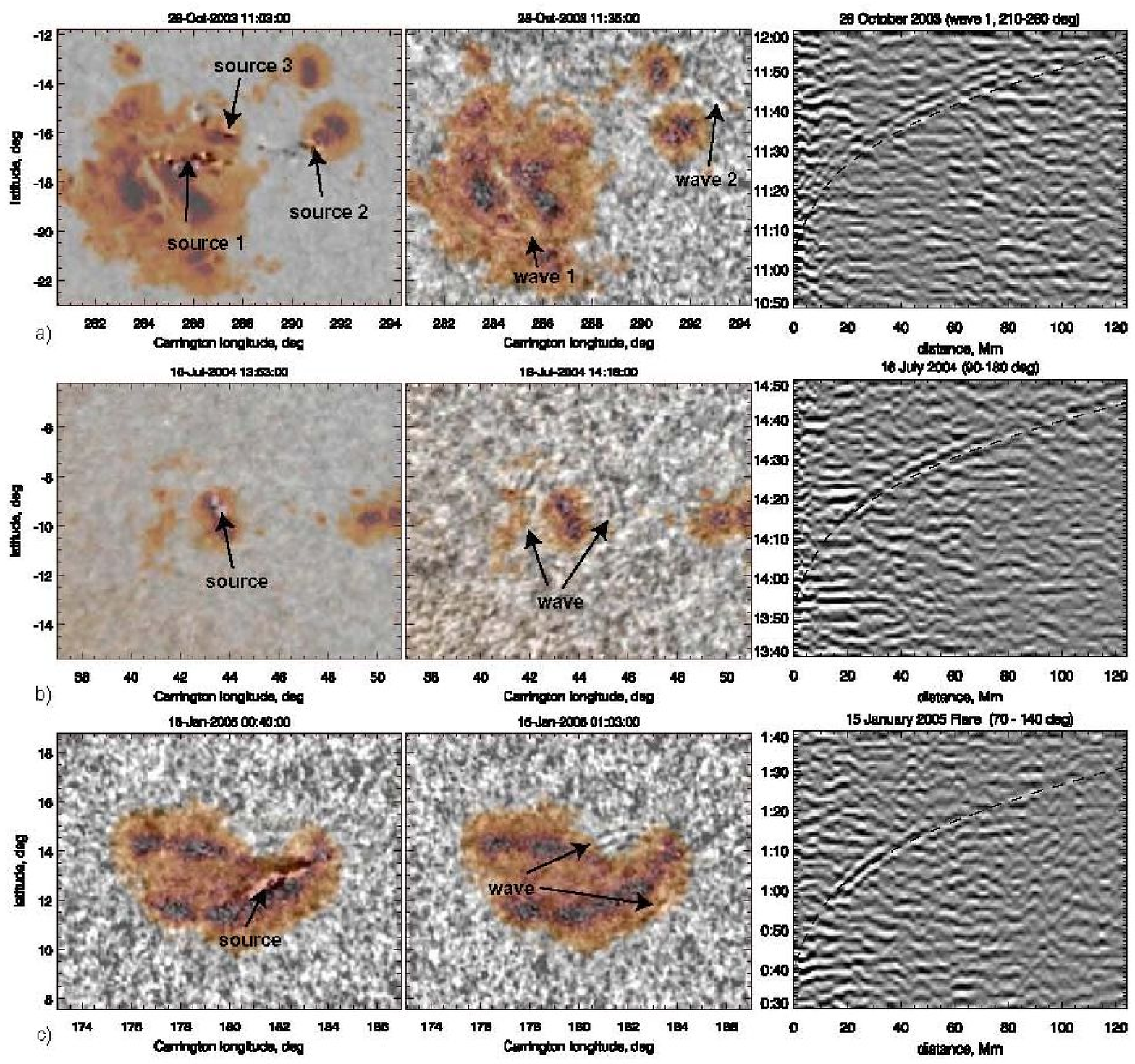}} \caption{
Observations of the seismic response of the Sun ('sunquakes") to
three solar flares: X17 of October 28, 2003 (top panels), X3 of July
16, 2004 (middle panels) and X1 flare of January 15, 2005. The left
panels show a superposition of MDI white-light images of the active
regions and locations of the sources of the seismic waves determined
from MDI Dopplergrams,  the middle column shows the seismic waves,
and the right panels show the time-distance diagrams of these
events. The dashed curve is a theoretical time-distance relation for
helioseismic waves.} \label{fig1}
\end{figure}

\clearpage

\begin{figure}
\centerline{\includegraphics[scale=0.7]{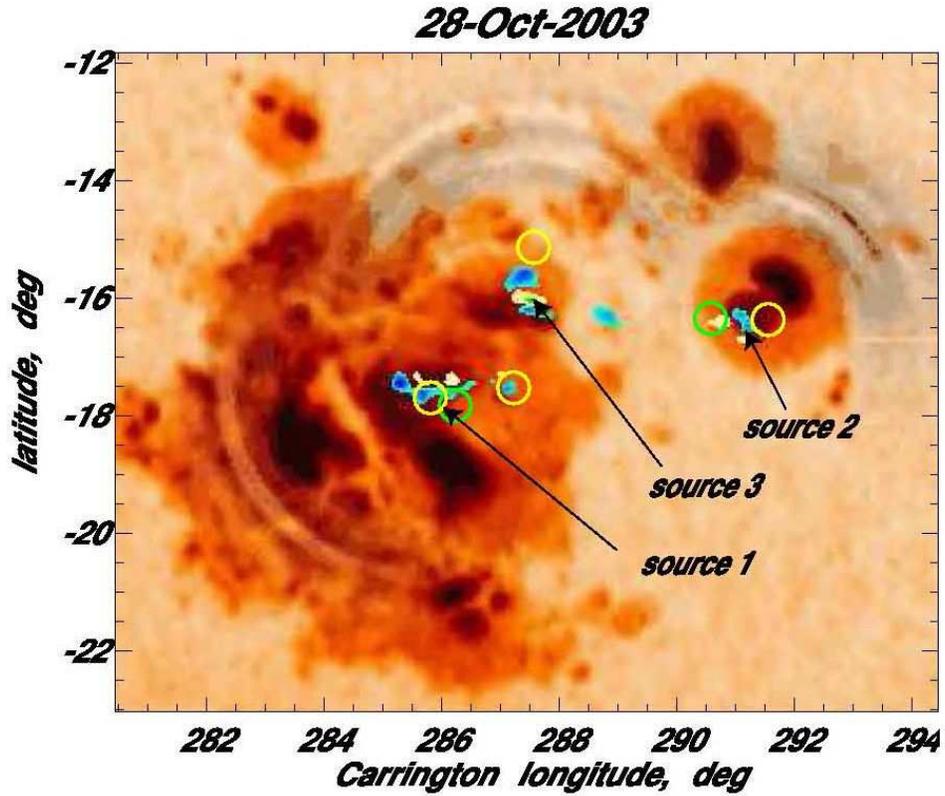}} \caption{. A
white-light image of active region NOAA 10696 observed on October
28, 2003, and superimposed images of the Doppler signal at the
impulsive phase, 11:06 UT, (blue and yellow spots show up and down
photospheric motions with variations in the MDI signal stronger than
1 km/s), positions of three wave fronts at 11:37 UT, and also
locations of the hard X-ray (50-100 keV) sources (yellow circles) at
11:06 UT, and 2.2 MeV gamma-ray sources (green circles). }
\label{fig2}
\end{figure}

\clearpage

\begin{figure}
\centerline{\includegraphics[scale=0.8]{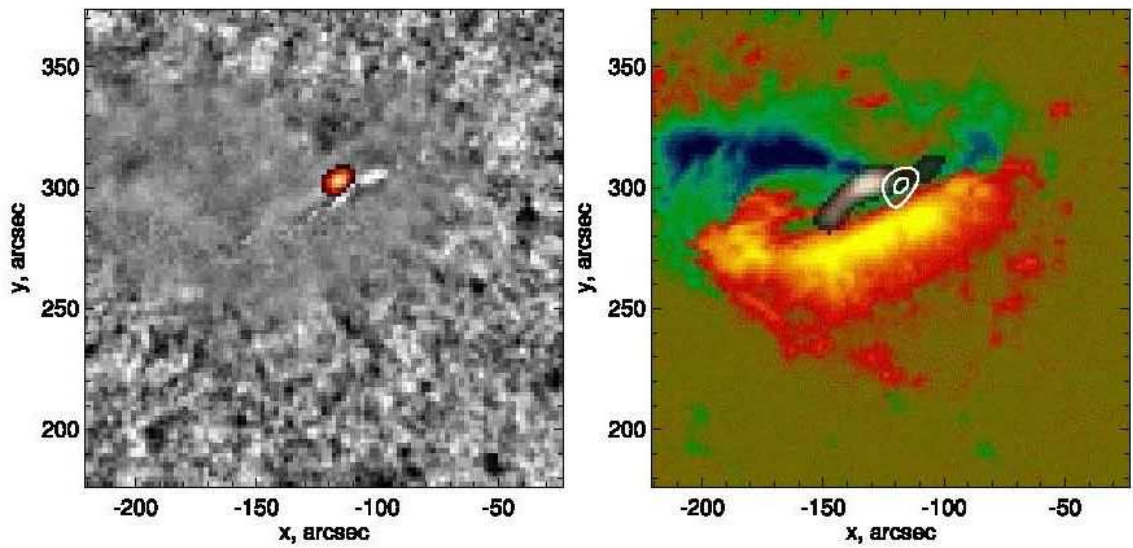}} \caption{ Seismic
and X-ray sources of the X1.2 flare of January 15, 2005. Left panel
shows the Dopplergram difference at 0:40 UT, in which the long white
feature near the center corresponds to strong downflows at the
seismic source and an image of the hard X-ray source (color spot).
The right panel shows the corresponding MDI magnetogram (red-
positive magnetic polarity of the line-of-sight component of
magnetic field, blue --negative polarity) and an image of the soft
X-ray emission (in gray) and contour line of the hard X-ray source.}
\label{fig3}
\end{figure}

\clearpage

\begin{figure}
\centerline{\includegraphics[scale=0.5]{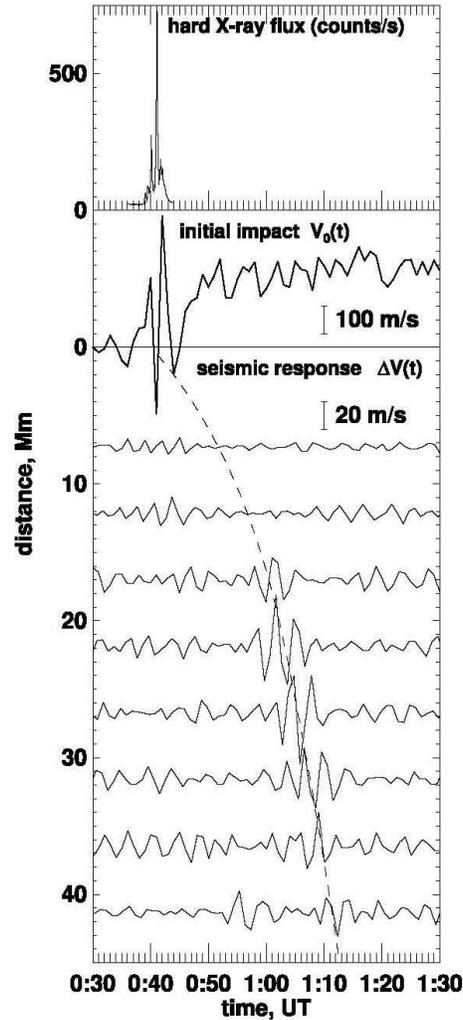}} \caption{. The
sequence of events during the flare of Janury 15, 2005. High-energy
electrons accelerated in the solar flare and interact with the lower
atmosphere, producing hard X-ray emission (observed by RHESS) and
shocks -- initial hydrodynamic impact in the photosphere (observed
by SOHO/MDI). Then, about 20 min after the initial impact an
expanding seismic wave was detected by SOHO/MDI. The dashed curve
shows a theoretical relation for helioseismic waves.} \label{fig4}
\end{figure}

\clearpage

\begin{table}[htbp]
\begin{center}
\caption{Solar flares with strong sunquake events.}
\begin{tabular}{|l|l|l|l|l|}
\hline
Date&
X-class&
Start time&
Peak time&
End time \\
\hline
9 July 1996&
X2.6&
09:01&
09:12&
09:49 \\
\hline
28 October 2003&
X17&
09:51&
11:10&
11:24 \\
\hline
29 October 2003&
X10&
20:37&
20:49&
21:01 \\
\hline
16 July 2004&
X3.6&
13:49&
13:55&
14:01 \\
\hline
15 August 2004&
M9.4&
12:34&
12:41&
12:43 \\
\hline
15 January 2005&
X1.2&
00:22&
00:43&
01:02 \\
\hline
\end{tabular}
\label{tab1}
\end{center}
\end{table}


\clearpage

\begin{thebibliography}{}

\bibitem[Donea et al.(1999)]{Donea1999} Donea, A.-C., Braun, D. C. {\&} Lindsey, C., 1999,
\apj, 513, L143

\bibitem[Donea \& Lindsey(2005)]{Donea2005} Donea, A.-C. {\&} Lindsey, C., 2005,
\apj, 630, 1168

\bibitem[Duvall et al.(1997)]{Duvall1997}Duvall, T. L., Jr. et al. 1997, Sol.
Phys., 170, 63

\bibitem[Fisher et al(1985)]{Fisher1985} Fisher, G. H., Canfield, R. C. \& McClymont, A. N., 1985, \apj, 289, 434

\bibitem[Hurford et al.(2004)]{Hurford2004} Hurford, G. et al. 2004,
in \textit{35th COSPAR Scientific Assembly}, 2402

\bibitem[Wolff(1972)]{Wolff1972} Wolff, C. L., 1972, \apj, 176, 833

\bibitem[Kosovichev(1986)]{Kosovichev1986} Kosovichev, A.G., 1986,
Bull. Crimean Astrophys. Obs., 75,  6

\bibitem[Kosovichev \& Zharkova(1995)]{Kosovichev1995} Kosovichev, A. G. {\&} Zharkova, V. V., 1995,
 in Helioseismology. Proc. 4th SOHO
Workshop, ESA SP-376, p.341


\bibitem[Kosovichev \& Zharkova(1998a)]{Kosovichev1998} Kosovichev, A. G. {\&} Zharkova, V. V., 1998,
Nature, 393, 317

\bibitem[Kosovichev \& Zharkova(1998b)]{Kosovichev1998a} Kosovichev,
A.~G., \& Zharkova, V.~V.\ 1998, IAU Symp.~185: New Eyes to See
Inside the Sun and Stars, 185, 191

\bibitem[Kosovichev et al.(2000)]{Kosovichev2000} Kosovichev, A. G., Duvall, T. L. J., Jr. {\&}
Scherrer, P. H., 2000, Sol. Phys., 192, 159

\bibitem[Kostiuk \& Pikelner(1975)]{Kostiuk1975} Kostiuk, N. D. {\&} Pikelner, S. B., 1975, Sov. Astr., 18, 590

\bibitem[Livshits et al.(1981)]{Livshits1981} Livshits, M. A., Badalian, O. G., Kosovichev, A. G. {\&} Katsova, M. M., 1981, Sol. Phys., 73, 269

%

\bibitem[Svestka(1970)]{Svestka1970} Svestka, Z., 1970, Sol. Phys., 13, 471
\end{thebibliography}
\end{document}